\documentclass[reprint,longbibliography,amsmath,amssymb,aps,nofootinbib]{revtex4-1}
\usepackage{multirow} %,bbm}
\usepackage{graphicx}% Include figure files
\usepackage{dcolumn}% Align table columns on decimal point
\usepackage{bm}% bold math
\usepackage{amssymb,amsmath}
\usepackage{tabularx}
\usepackage{subcaption}
\usepackage{flushend}% align the last two columns
\usepackage{lipsum}% align the last two columns: better add \onecolumngrid after refs
\usepackage{float}
\usepackage{amsmath,multirow,bigdelim}
\usepackage{contour}
\usepackage{array}
\usepackage[bottom]{footmisc}%footnote at the bottom of the page
\usepackage{soul}%strike out words \st{}
\usepackage{hyperref}%include url
\newcolumntype{P}[1]{>{\centering\arraybackslash}p{#1}}
\usepackage{rotating} % <-- HERE
\usepackage{slashed}
\usepackage[normalem]{ulem}

\usepackage{amssymb}% http://ctan.org/pkg/amssymb
\usepackage{pifont}% http://ctan.org/pkg/pifont
\newcommand{\beq}{\begin{equation}}
\newcommand{\eeq}{\end{equation}}
\newcommand{\bea}{\begin{eqnarray}}
\newcommand{\eea}{\end{eqnarray}}
\newcommand{\ba}{\begin{array}} 
\newcommand{\ea}{\end{array}}

\newcommand{\ii}{\boldsymbol{i}}

\newcommand\gev{\,\mathrm{GeV}}
\newcommand\tev{\,\mathrm{TeV}}
\newcommand\eg{{\it e.g.}}
\newcommand\ie{{\it i.e.}}
\newcommand{\gsim}{\lower.7ex\hbox{$\;\stackrel{\textstyle>}{\sim}\;$}}
\newcommand{\lsim}{\lower.7ex\hbox{$\;\stackrel{\textstyle<}{\sim}\;$}}
\newcommand{\invab}{\,\mbox{ab}^{-1}\!}

\definecolor{red}{rgb}{.9,0,0}
\definecolor{green}{rgb}{0,0.7,0}
\definecolor{blue}{rgb}{0,0,0.9}
\definecolor{orange}{rgb}{0.8,0.3,0}

\everymath{\displaystyle}

\begin{document}

%\title{Updated limits on $\lambda'_{ijk}$ couplings in the R-parity-violating MSSM}
\title{Constraining R-parity-violating couplings in $\tau$-processes at the LHC and\\ in electroweak precision measurements}

\author{{Saurabh Bansal}$^a$}% \email{sbansal@nd.edu} 
\author{{Antonio Delgado}$^a$}% \email{rcapdevi@nd.edu}
\author{{Christopher Kolda}$^a$}% \email{ckolda@nd.edu}
\author{{Mariano Quiros}$^{a,b}$}% \email{quiros@ifae.es}
\affiliation{$^a$ Department of Physics, University of Notre Dame, 225 Nieuwland Hall, Notre Dame, Indiana 46556, USA\\ }

\affiliation{$^b$  Institut de F\'{\i}sica d'Altes Energies (IFAE) and BIST, Campus UAB\\ 08193, Bellaterra, Barcelona, Spain
}

\begin{abstract}
We find new limits on the $\lambda'_{3jk}$ R-parity violating (RPV) couplings of the minimal supersymmetric standard model, using Drell-Yan differential cross sections at the LHC and electroweak precision measurements from LEP and SLC. Specifically, limits on six out of the nine $\lambda'_{3jk}$-couplings, with $j=~1~\text{or}~2$, are obtained using Drell-Yan data, with the remaining three (for $j=3$) bounded by precision electroweak data. We also update the limits on all $\lambda'_{ijk}$-couplings using electroweak data and find new bounds on $\lambda'_{132}$ and $\lambda'_{232}$ that are stronger than obtained elsewhere.
A table of all current bounds on $\lambda'_{ijk}$ is given in an appendix.
\end{abstract}

\maketitle

%% ===================================================== %%
%% =================== FIRST SECTION ===================== %%
%% ===================================================== %%
\section{\label{section1}Introduction}

With the discovery of the Higgs boson and the completion of the Standard Model (SM) particle spectrum, the search for physics beyond the Standard Model (BSM) is well underway.  Among the wide range of possible BSM models, supersymmetry (SUSY) offers a particularly attractive alternative, providing a solution to the hierarchy problem and offering a robust framework for approaching many of the other problems of the SM. But while SUSY may have some very strong theoretical motivations, experiments have yet to find signatures of the most minimal SUSY model in its expected parameter space. This has led physicists to explore less minimal versions of SUSY, including its R-Parity Violating version. 

R-parity was originally imposed on the Minimal Supersymmetric Standard Model (MSSM) in an attempt to avoid problems with fast proton decay; as a side effect, it generated an attractive dark matter candidate in the form of the lightest (necessarily stable) SUSY particle (LSP). But the presence of R-parity (or lack thereof) also plays a key role in determining how one should search for the presence of SUSY at colliders like the LHC. In particular, models with unbroken R-parity always pair-produce SUSY particles, and leave missing energy signatures as the SUSY particles decay down to the LSPs, which escape the detector unseen. Most of the key strategies for studying SUSY at colliders involve this particular production path: on-shell pair production, leading to missing energy in the detector. In turn, very strong bounds have been placed on the majority of SUSY particles, constraining their masses to be generally above $500\gev$ to $2\tev$~\cite{Canepa:2019hph}.

In SUSY models with R-parity violation (RPV), the above search strategy generally fails, though other search avenues open up that can help fill the breach. In a previous work (Ref.~\cite{Bansal:2018dge}), we studied a technique for placing strong constraints on the parameter space from the study of Drell-Yan (DY) processes (both neutral and charged current) at the LHC. In models with RPV, SUSY partners can be exchanged by SM particles at tree level, something that is not possible in R-parity conserving models, leading to sizable interference effects in SM processes. In Ref.~\cite{Bansal:2018dge}, we analyzed the effect of an RPV coupling $\lambda'_{ijk}$ (there are 27 such couplings, defined in the next section) on DY processes involving electrons and muons final states. We found that the LHC could already place surprisingly strong constraints in wide regions of the parameter space of squark masses and $\lambda'_{ijk}$, for couplings to the first- and second-generation leptons. 

In this paper we return to this subject and work to place bounds for the case in which the only available RPV coupling is to $\tau$-leptons using the same techniques. In addition, we revisit precision electroweak constraints on RPV, mostly derived from the LEP and SLC data, to show that the current best fits place stronger constraints on certain $\lambda'_{ijk}$ couplings than previously advertised. In all, we will present new bounds on 11 of the 27 $\lambda'_{ijk}$ couplings, all of which are stronger than existing bounds found in the literature.

The paper is organized as follows: in section~\ref{section2} we will introduce the model and processes that will be studied; we will then present the analysis and results for Drell-Yan processes in section~\ref{section3} and electroweak processes in~\ref{section4}; we will devote section~\ref{section5} to our conclusions. For completeness, we summarize in the appendix the current bounds on all the 27 $\lambda'_{ijk}$ couplings.

%% ===================================================== %%
%% =================== SECOND SECTION ==================== %%
%% ===================================================== %%
\section{\label{section2}$L$-violating RPV }

R-parity is a multiplicative quantum number, defined as
\begin{equation}
R_P=(-1)^{3(B-L)+2s} \,,
\label{eq:RP}
\end{equation}
where $B$ is the baryon number, $L$ the lepton number and $s$ the spin of a specific state. R-parity is usually enforced in the MSSM and allows, as the most general superpotential,
$$\mathcal W_{R_P}=Y^u_{ij}U^c_i Q_j H_u-Y^d_{ij}
D^c_i Q_j H_d-Y^e_{ij}E^c_i L_j H_d+\mu H_u H_d
$$
where $H_u$ and $H_d$ are two Higgs doublets with hypercharges $\pm1/2$, respectively. Here, $L$ and $Q$ are the $SU(2)$ doublets, while $E^c$, $D^c$ and $U^c$ are singlets, and $Y^{U}_{ij}$, $Y^{D}_{ij}$ and $Y^{E}_{ij}$ the Yukawa coupling matrices. The above superpotential %(\ref{eq:WRP}) 
prevents tree-level processes where $B$ and/or $L$ are violated and has strong phenomenological implications, including the fact that supersymmetric partners should be pair produced in colliders and that the LSP is stable and can become a dark matter candidate. However, in view of the strong experimental bounds on theories with R-parity conservation, one can extend the above superpotential by including R-parity-breaking terms, which soften the present experimental bounds.

In the MSSM, the RPV portion of the trilinear superpotential can be written as, 
$$
\mathcal{W}_{\slash \hspace{-0.2cm}R_P} =\frac{1}{2} \lambda_{ijk} L_i L_j E^c_k+\lambda'_{ijk} L_i Q_j D^c_k+\frac{1}{2}\lambda''_{ijk} U^c_i D^c_j D^c_k.
$$
Using the standard notation, we have defined $\lambda_{ijk}$, $\lambda'_{ijk}$ and $\lambda''_{ijk}$ as new Yukawa couplings, where $i$, $j$ and $k$ are the generation indices; we omit a bilinear term that mixes sleptons and Higgs fields. In order to enforce $B$-conservation, we assume the $\lambda''_{ijk}$ are all zero, but $\lambda_{ijk}$ and $\lambda'_{ijk}$ remain. Moreover, for the purpose of this work, we will only concentrate on the $\lambda'_{ijk}$ interactions.  
Furthermore, as we did in Ref.~\cite{Bansal:2018dge}, we will only take one element of $\lambda'_{ijk}$ at a time to be non-zero, in order to avoid the (possibly) complicated interference effects, and the potentially large contributions to quark or lepton flavor-changing amplitudes.

As we mentioned above, in Ref.~\cite{Bansal:2018dge}, the constraints from Drell-Yan processes on $\lambda'_{ijk}$ couplings involving electrons and muons, \ie~$i=1\text{ and }2$, were studied. In this paper, we will concentrate on $\lambda'_{ijk}$ couplings involving taus, \ie~$i=3$. In this scenario, the RPV superpotential leads to the following Lagrangian in the $up$-quark mass basis,
\bea
\mathcal{L}& = \lambda'_{3jk} \Big[ \Big((V\bar{d}^c)_j P_L \nu_\tau - \bar{u}^c_j P_L \tau \Big) \tilde{d}_{Rk}^* + \Big(\bar d_k P_L \nu_\tau (V\tilde{d}_L)_j-\nonumber\\
\label{RPVlag}& \bar d_k P_L \tau \tilde{u}_{Lj}\Big)
+\left(\bar d_k P_L (V d)_j \tilde{\nu}_{\tau L}+\bar d_k P_L u_j \tilde{\tau}_L\right)\Big]+ h.c.
\label{Lagrangian}
\eea 
Here $V$ is the CKM matrix, which we will consider in the (reasonable for our purposes) diagonal approximation. 

One of the key points to notice is that the $\lambda'_{ijk}$ couplings cause the squarks to couple as scalar leptoquarks; that is, they interact with both quarks and leptons at a single vertex. As such, squarks can be exchanged in the $t$-channel in DY scattering and can appear in loops in $Z$-decay, two processes that we will consider in the following two sections.

~

In the next two sections, we present our analysis and results for Drell-Yan and electroweak processes, separately.

\section{\label{section3}Constraints from Drell-Yan processes}

The contributions of scalar leptoquarks to DY scattering processes were recently studied in both the neutral current channel~\cite{Raj:2016aky,Alves:2018krf} and the charged current channel~\cite{Bansal:2018eha}. As with these previous papers, we will refer to the neutral current case as dilepton DY scattering, since the final state is always $\ell^+\ell^-$ with both leptons of the same flavor (assuming only a single RPV coupling at a time), and the charged current as monolepton DY scattering, since the final state is of the form $\ell\nu_\ell$. In Ref~\cite{Bansal:2018dge}, we specialized this analysis to RPV SUSY, studying the $\lambda'_{ijk} L_iQ_jD^c_k$ superpotential coupling, for $i=1$ (final state electrons) and $i=2$ (final state muons). In both of these final states, excellent analyses of DY data had been completed by both ATLAS and CMS at $\sqrt{s} = 13\tev$. 

The situation for final state $\tau$-leptons is not as simple at present. DY mono-tau searches have been published by both ATLAS~\cite{Aaboud:2018vgh} and  CMS~\cite{Sirunyan:2018lbg} over a wide range of transverse mass. But a similar search for the di-tau process is not available from either of the collaborations. There is a measurement of di-taus from the CMS collaboration~\cite{Sirunyan:2018qio}, but it is limited to small invariant masses ($m_{\tau\tau}<250\gev$) and low integrated luminosity ($2.3\,\text{fb}^{-1}$). The low invariant mass range studied, and the large statistical errors, will prevent us from placing strong constraints using that data set. Due to these limitations, we will only be able to use the mono-tau DY process to constrain the RPV couplings involving $\tau$-leptons.

\begin{figure}[!t]
	\begin{subfigure}[b]{0.235\textwidth}
		\includegraphics{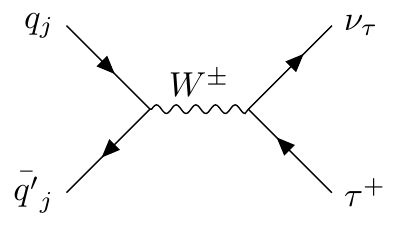} 
	\end{subfigure}
	\begin{subfigure}[b]{0.235\textwidth}
		\includegraphics{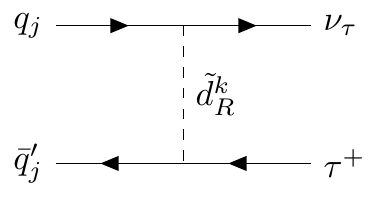} 
	\end{subfigure}
	\caption{\small \label{fig:FeynDiagMono}Feynman diagrams for mono-tau production in the SM (left) and supersymmetry with RPV (right).}
\end{figure}

The $\lambda'_{ijk}$ couplings contribute to the mono-tau process due to the first two terms in Eq.~(\ref{Lagrangian}) involving $\tilde d_R^k$. After integrating out $\tilde d_R^k$, these terms generate the operator $(\bar \nu_\tau P_R d^c_j)(\bar u^c_j P_L \tau)$, which, after fierzing, equals $(1/2)(\bar d_j \gamma^\mu P_L u_j)(\bar\nu_\tau \gamma_\mu P_L \tau)$. Thus, this operator directly interferes with the SM process. The Feynman diagrams for this process in the SM and RPV SUSY are shown in Figure~\ref{fig:FeynDiagMono}, with the latter contributing through a $\tilde d_R^k$-mediated $t$-channel process with $d_j$ and $u_j$ in the initial state. Note that after integrating out $\tilde d_R^k$, the contribution from RPV SUSY only depends on the value of $j$. In other words, for a fixed value of $j$, the constraints on $\lambda'_{3jk}$ are the same for all $k$.

Since quarks in the initial state at LHC have differing parton distribution functions (PDFs), the constraints on $\lambda'_{3jk}$ strongly depend on the value of $j$. The constraints are strongest for $j=1$, due to the large PDFs of $u$- and $d$-quarks, followed by somewhat weaker constraints for $j=2$, weaker mostly because of the suppressed PDFs of $c$-quarks. But since the PDF of the top quark can be taken to be zero, $\lambda'_{3jk}$ with $j=3$ is not constrained at all. In total, the mono-tau analysis using LHC data can be used to constrain six out of ten $\lambda'_{3jk}$ couplings.

We now provide a brief overview of our analysis for Drell-Yan processes, and refer the interested reader to Ref.~\cite{Bansal:2018dge} for a more detailed account. The analysis done here is similar to that of Ref.~\cite{Bansal:2018dge}, with final state electrons and muons replaced by taus. As mentioned above, due to the lack of a di-tau search at the LHC in the right kinematic regime and with sufficient luminosity, we will not be able to use di-tau data to constrain our parameter space. But when the required analysis has been performed by ATLAS and/or CMS, the work done here can be replicated for di-tau processes as well, though a preferable option would be for the collaborations to complete their own interpretation of DY data in terms of bounds on RPV SUSY couplings, following the model here and in Ref.~\cite{Bansal:2018dge}.

As discussed above, the mono-tau DY bounds on $\lambda'_{3jk}$ are independent of $k$; that is, the squark mass bounds we obtain are the same for each of the $\tilde d_R^k$ squarks, for $k=1,2,3$. And since we always take $i=3$ (final state as $\tau$-leptons), the only remaining dependence is on $j$. For each value of $j$, we calculate the constraints in the $m_{\tilde d_R^k}~\text{versus}~\lambda'_{3jk}$ plane, by comparing the RPV prediction to the experimental measurements. These measurements are published by both ATLAS~\cite{Aaboud:2018vgh} and  CMS~\cite{Sirunyan:2018lbg} collaborations at $\sqrt{s} = 13\tev$ with 36~fb$^{-1}$ of integrated luminosity. In this work, we compare with the ATLAS data as they are readily available at \url{ www.hepdata.net}. The signal events are obtained by simultaneously calculating the SM plus RPV contribution to the transverse mass $(m_T)$ spectrum for $pp\to\tau\nu_\tau$ processes. These calculations are done analytically at the leading order, using the MSTW 2008 NNLO PDFs~\cite{Martin:2009iq}. The resulting spectrum is then rescaled bin-by-bin to account for the higher-order corrections and lepton reconstruction efficiency, so that our SM (``background") spectrum matches the irreducible background data from ATLAS. Finally, the net signal plus background is obtained by adding the reducible background to the generated event distribution.

\begin{figure}[!tbh]
		\includegraphics[width=.95\linewidth]{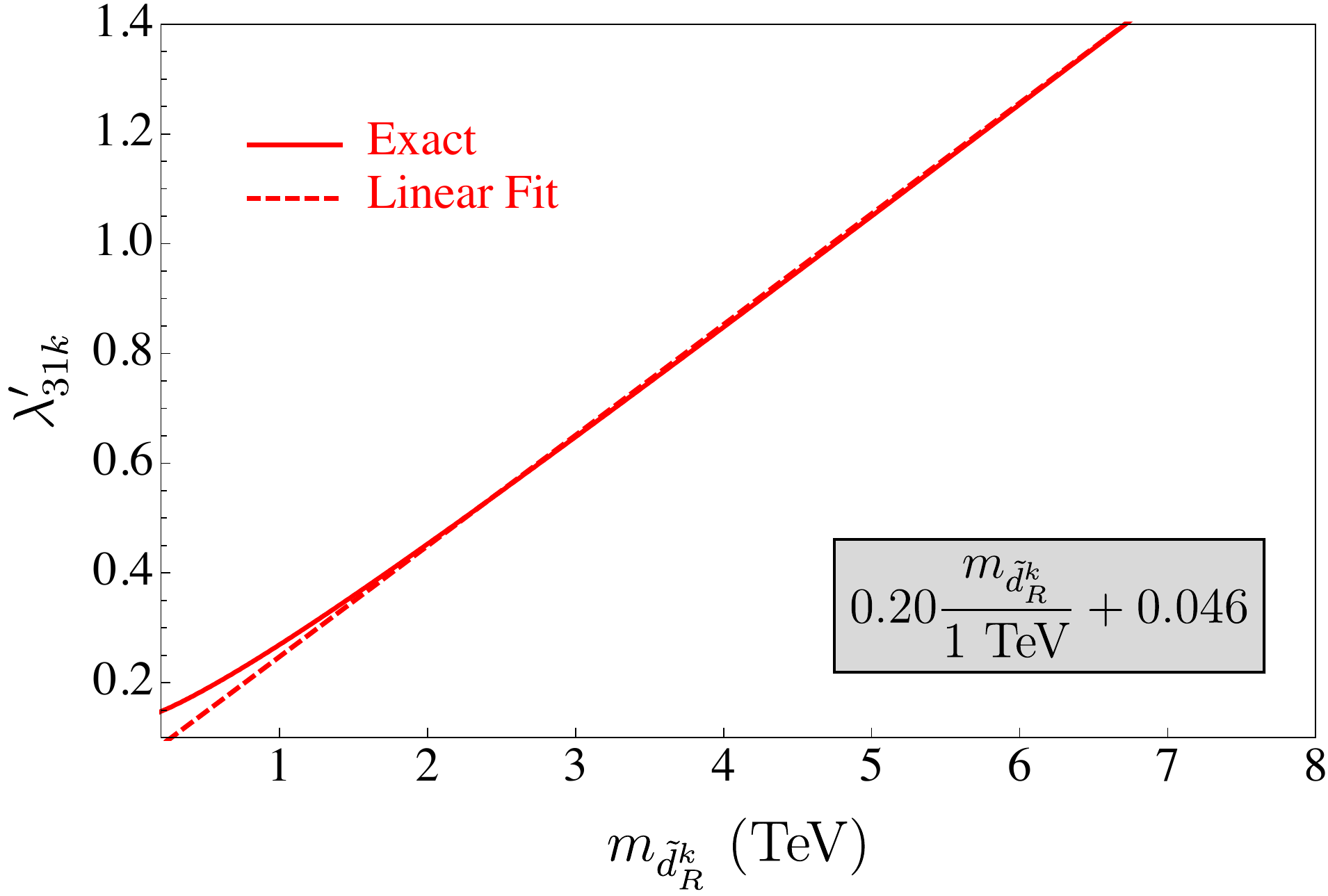} 
	\caption{\small \label{fig:MonoExample}The 95\% C.L.\ upper bound (solid line) on $\lambda'_{31k}$ using monolepton data. A linear fit (dashed line) to this bound  and the corresponding equation are also shown.}
\end{figure}

To quantify the effect of our signal, and to estimate the limits on the RPV parameters, we use a conservative version of a $\Delta\chi^2$ test. Specifically, 95\% C.L. limits are obtained when $\Delta\chi^2\equiv\chi^2_\text{model}-\chi^2_\text{SM} = 5.99$. The systematic errors used in this test are taken, bin by bin, from the ATLAS searches~\cite{Aaboud:2018vgh}, which range from about 15\% at low transverse masses ($\sim 200 $ GeV) to over 50\% at high transverse masses. Finally, we quote our limits on the model by fitting a straight line in the $(m_{\tilde d_R^k},\lambda'_{3jk})$-plane to the 95\% C.L. contour in the region where $m_{\tilde d_R^k}\geq 1\tev$. Below $1\tev$ squark masses, the constraints are far less linear, and the squarks themselves are often better constrained by direct production limits. Thus, the limits we quote only hold for $m_{\tilde{d}_R^k}>1\tev$. The accuracy of our linear fit can be seen in Figure~\ref{fig:MonoExample}, where the 95\% C.L. contour (solid line) and a linear fit (dashed line) to this contour are shown for $\lambda'_{31k}$. %For lower squark masses, the bounds on $\lambda'_{ijk}$ depend on the squark mass in a more complicated way. However, for such light squarks one must also contend with strong direct production constraints.

Our resulting constraints from monolepton DY processes are summarized in the second column of Table~\ref{tab:i1}. The first column indicates the existing bound in the literature, collected and updated in Refs.~\cite{Barger:1989rk,ledroit:in2p3,Allanach:1999ic,Barbier:2004ez,Chemtob:2004xr,Dercks:2017lfq}. These bounds are derived from: $R_{\tau\pi}=\Gamma(\tau\to\pi\nu_\tau)/\Gamma(\mu\to\pi\nu_\mu)$ for $\lambda'_{31k}$; $R_{D_s}=\Gamma(D_s\to\tau\nu_\tau)/\Gamma(D_s\to\mu\nu_\mu)$ for $\lambda'_{32k}$; and $R_\tau=\Gamma(Z\to \text{had})/\Gamma(Z\to\tau\bar\tau)$ for $\lambda'_{33k}$. The constraints on $\lambda'_{33k}$ are discussed in more detail below.

\begin{table*}[htb]
	\begin{ruledtabular}
		{\renewcommand{\arraystretch}{2.5}
			\begin{tabular}{||c |c c c c||}
				$ijk$ %of $\lambda'_{ijk}$
				&
				Literature & Monolepton & Electroweak & 
				Projected \\
				\hline \hline
				$31k$ & $1.1  \frac{m_{\tilde{d}_R^k}}{1 \tev}$ & \fbox{$0.20 \frac{m_{\tilde{d}_R^k}}{1 \tev}+0.046$} & $1.5 \frac{m_{\tilde{f}}}{1 \tev}+0.41$ & $0.14 \frac{m_{\tilde{d}_R^k}}{1 \tev}+0.046$ \\

				$32k$ & $5.2  \frac{m_{\tilde{d}_R^k}}{1 \tev}$ & \fbox{$1.5 \frac{m_{\tilde{d}_R^k}}{1 \tev}+0.66$} & $1.5 \frac{m_{\tilde{f}}}{1 \tev}+0.41$ & $0.75 \frac{m_{\tilde{d}_R^k}}{1 \tev}+0.69$ \\ 
				
				$33k$ & $1.24^\dagger$ & $-$ & \fbox{$0.54 \frac{m_{\tilde{f}}}{1 \tev}+0.38$} & $-$ 
				
			\end{tabular}
		}
	\end{ruledtabular}
	\caption{\small \label{tab:i1}%
		Upper bounds on $\lambda'_{3jk}\ (k=1,2,3)$ from the literature and derived in this study. The strongest current constraint on a particular coupling (for $m_{\tilde{f}} > 1 \tev$) is shown in a box. Projected bounds are obtained using a mono-tau analysis assuming $3\invab$ of data.~~$^\dagger$See text for an explanation of this bound which holds for sfermion masses of 1 TeV.}
\end{table*}

One must note that the current ATLAS mono-tau data show a small {\it excess}\/ over the SM predictions for most of the $m_T$-bins. On the other hand, the RPV monolepton operator interferes {\it destructively}\/ with the SM, pulling down the expected cross section. Thus the resulting constraints on $\lambda'_{3jk}$ are much stronger than one would expect just by comparing to the SM distribution. This same excess of events in the data also results in a constraint on $\lambda'_{32k}$ that is surprisingly strong despite being suppressed by the $c$-quark PDF. Thus one should keep in mind that if the mono-tau data were to become more closely aligned with the SM prediction, the bounds would weaken. Similar observations were also made for mono-muons in Ref.~\cite{Bansal:2018dge}.
 
Unsurprisingly, the limits we obtain are very sensitive to current experimental uncertainties in the DY spectrum, and so these limits may strengthen or weaken with higher integrated luminosity, at least at first. But we can also make a simple projection of the expected limits on $\{m_{\tilde q}, \lambda'_{ijk}\}$ at the High Luminosity LHC with $3\invab$ of integrated luminosity and $\sqrt{s} = 13\tev$, assuming the data matches the SM predictions. In this analysis, we again use the transverse mass dependent systematic error as seen across bins in the current ATLAS data and neglect all sources of reducible background. The last column of Table~\ref{tab:i1} indicates these expected limits from a $3\invab$ high luminosity LHC. 

As expected, the constraints obtained from the LHC monolepton analysis are strongest when the quarks involved are first generation, and weaken significantly when one requires second generation quarks in the initial state (by roughly a factor of 10 in going from $\lambda'_{31k}$ to $\lambda'_{32k}$). In such a case, the dilepton bound would be expected to outdo the monolepton bound, which is another reason why the absence of an experimental DY dilepton analysis hampers the setting of the strongest possible bounds right now. For $\lambda'_{33k}$, monolepton searches at the LHC are completely insensitive and another route must be sought in order to obtain bounds on this coupling; we discuss this case in the next section.

\section{\label{section4} Electroweak Constraints}

\begin{figure}[!]	
	\centering
	\begin{subfigure}[b]{0.23\textwidth}
		\includegraphics{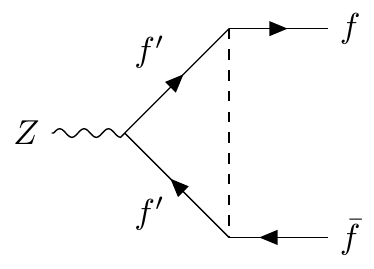} 
		\caption{}\label{FeynDiagEW1}
	\end{subfigure}
	\begin{subfigure}[b]{0.23\textwidth}
		\includegraphics{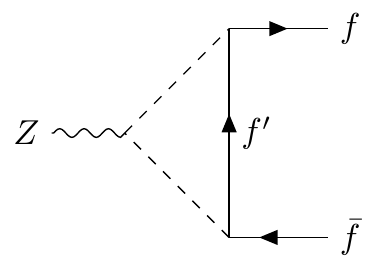} 
		\caption{}\label{FeynDiagEW2}
	\end{subfigure}
	\begin{subfigure}[b]{0.23\textwidth}
		\includegraphics{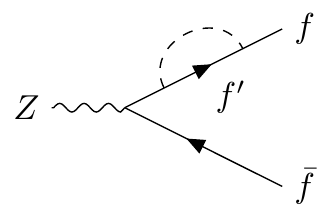} 
		\caption{}\label{FeynDiagEW3}
	\end{subfigure}
	\begin{subfigure}[b]{0.23\textwidth}
		\includegraphics{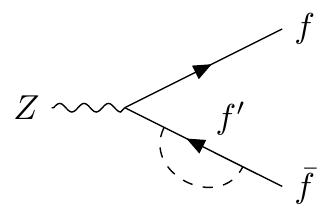} 
		\caption{}\label{FeynDiagEW4}
	\end{subfigure}
	\caption{\small\label{fig:FeynDiagEW}Feynman diagrams for $Z\to f\bar f$ via RPV couplings. The dashed lines represent the exchange of the scalars $S$ and/or $\tilde S$ defined in the text.}
\end{figure}

LHC bounds on the $\lambda'_{33k}$ couplings are non-existent currently, and will be weak, at best, in the future (using dilepton data, once available). Luckily, there is a well-known auxiliary bound that can be placed on all of the RPV couplings coming from precision electroweak observables. 
In particular, the RPV couplings can affect electroweak observables at one loop by modifying the coupling strength of $Z$ to fermions. Since these observables have been measured very precisely by both LEP and SLC at, and above, the $Z$-pole, the $\lambda'_{ijk}$ couplings can be constrained by quantifying these modifications. This requires the calculation of RPV-induced $Z\to f \bar f$ one-loop diagrams, shown in Figure~\ref{fig:FeynDiagEW}, where $f$ is any of $\ell_i$, $\nu_i$, $u_j$, $d_j$ and $d_k$ for a non-zero $\lambda'_{ijk}$. 
Each final state $f$ can have multiple different fermions and sfermions in the loop; for example, a non-zero $\lambda'_{332}$ can lead to $Z\to \tau \bar \tau$ with $t_L$-quark and $\tilde s_R$ as well as $s_R$-quark and $\tilde t_L$ in the loop. By calculating all the possible one-loop diagrams for a single non-zero $\lambda'_{ijk}$, the RPV parameter space is constrained.

One may expect that the effect of one-loop diagrams with $\mathcal{O}(\tev)$ sfermion masses would be negligible at the $Z$-pole. But it has long been known (\eg, \cite{Mizukoshi:1994zy,ColuccioLeskow:2016dox,Bansal:2018nwp}) that the contribution of these diagrams with $t$-quarks in the loop can be fairly large. This is the result of helicity flips on the $t$-quark lines. Since the $\lambda'_{ijk}$ couplings can contribute a $t$-quark for $j=3$, we expect relatively strong electroweak constraints on $\lambda'_{i3k}$. In this section, we will analyze the contributions of RPV to precision electroweak observables in order to constrain the set of couplings $\lambda'_{i3k}$. We are hardly the first to do this analysis; previous analyses were performed by Refs.~\cite{Godbole:1992fb,Bhattacharyya:1995pr,Yang:1999ms,Lebedev:1999vc,Lebedev:1999ze,ledroit:in2p3}. Our analysis uses updated electroweak parameters (mostly updated SM predictions for the precision electroweak observables) that have the effect of strengthening the bounds somewhat. But we will also be using an alternative observable, defined below, that allows for an improvement on the electroweak constraints on $\lambda'_{33k}$ of roughly 25\% for sfermion masses of $1\tev$. As a cross check, we also performed our fits using the observables, data and SM fits of previous authors, and successfully reproduced the previously obtained bounds.

Before specializing to the problem at hand, we first stop to consider a somewhat more general calculation. Consider a toy model with two scalars, $S$ and $\tilde S$, which interact with the fermions of the SM through the Lagrangian:
\begin{align}
\label{TOYlag}
\mathcal{L} = \sum_{X=L,R} \left\{ \lambda_X S\, \overline{f'^c} f_X+\tilde \lambda_X  \tilde{S} \bar{f'}  f_X + h.c.\right\}
\end{align}
Here $f_{L,R}$ and $f'_{L,R}$ are chiral SM states with their usual gauge quantum numbers, while the quantum numbers of $S$ and $\tilde S$ are chosen such that the terms in the Lagrangian are gauge invariant.
These scalars can then modify the $Z\to\bar f f$ amplitude through the Feynman diagrams shown in Fig.~\ref{fig:FeynDiagEW}, with $f'$ being the loop fermion. 
Since there is a discontinuity in the amplitude of the first diagram (Fig.~\ref{FeynDiagEW1}) at $m_{f'} = m_Z/2$, the contributions from these diagrams can be divided into two cases: for $m_{f'}\simeq 0$ and $m_{f'}>m_Z/2$. All the fermions of the SM fall into the first category except for the $t$-quark, which clearly belongs to the second case. Meanwhile, we assume the common mass of all scalars, denoted by $M$, to be much greater than $m_Z$ and thus there is no discontinuity in the second diagram (Fig.~\ref{FeynDiagEW2}).

We can absorb the contributions from the one-loop diagrams into a redefinition of the the $Z\bar f f$ couplings, via:
\begin{equation}
\mathcal L_Z=\frac{g_2}{c_w}\,\sum_{X=L,R}\left( g_{X}^f+\Delta g_{X}^f \right)Z_\mu\, \bar f_{X}\gamma^\mu f_{X}
\end{equation}
where $g_L^{f} = T_3^{f}-Q^{f} s_w^2$, $g_R^{f}=-Q^{f} s_w^2$, and  $T_3^{f}$ is the weak $T_3$ of the left-handed fermion $f_L$.
The leading order corrections to the $Z\bar f f$ couplings due to $S$, taken from Appendix B of Ref.~\cite{Bansal:2018nwp}, can be written as:

~

\noindent {\bf Case I:} $m_{f'}\sim 0$: 
\begin{align}\label{Appdeltagmz1}
&\Delta g^{f}_{L/R}  =\frac{|\lambda_{L/R}|^2 m_Z^2 N_c^{f'}}{288 \pi^2 M^2 N_c^f} \times \nonumber \\
&~~~~~~~ \left\{g^f_{L/R}-g^{f'}_{L/R}\left(12 \log \frac{M}{m_Z}+1+\ii 6 \pi\right)\right\}
\end{align}

\noindent {\bf Case II:} $m_{f'}> m_Z/2$:
\begin{align}\label{AppdeltagmL1}
&\Delta g^f_{L/R}  = \pm \frac{ |\lambda_{L/R}|^2 m_{f'}^2 T_3^{f'} N_c^{f'}}{16 \pi^2  M^2 N_c^f} \left(2 \log \frac{M}{m_{f'}}-1\right) +\nonumber \\
&\frac{|\lambda_{L/R}|^2 m_Z^2 N_c^{f'}}{288 \pi^2 M^2 N_c^f} \left\{g^f_{L/R}-g^{f'}_{L/R}\left(12 \log \frac{M}{m_{f'}}-9\right)\pm 3 T_3^{f'}\right\},
\end{align}
where $M$ is the mass of $S$, and 
$N_c^{f,f'}$ is the color factor of the final fermion state, $f$, or of the internal loop fermion, $f'$ (\ie, 3 for a color triplet, 1 for a singlet). 
The imaginary piece that arises in Case~I is due to the loop fermions going on shell, but does not contribute to $Z$-pole observables. Also note that the $\pm 3T_3^{f'}$ term at the end of Eq.~(\ref{AppdeltagmL1}) corrects a typographical error in Ref.~\cite{Bansal:2018nwp}.

Similarily, on computing the amplitudes involving $\tilde S$, we find the dominant parts of the corrections to $Z\bar f f$ couplings for the two cases to be:

~

\noindent {\bf Case I:} $m_{f'}\sim 0$: 
\begin{align}\label{Appdeltagmz2}
&\Delta g^{f}_{L/R}  =\frac{|\tilde\lambda_{L/R}|^2 m_Z^2 N_c^{f'}}{288 \pi^2  M^2 N_c^f} \times \nonumber \\
&~~~~~~~ \left\{g^f_{L/R} + g^{f'}_{R/L}\left(12 \log \frac{M}{m_Z}+1+\ii 6 \pi\right)\right\}
\end{align}

~

\noindent {\bf Case II:} $m_{f'}> m_Z/2$:
\begin{align}\label{AppdeltagmL2}
&\Delta g^f_{L/R}  = \pm \frac{ |\tilde \lambda_{L/R}|^2 m_{f'}^2 T_3^{f'} N_c^{f'}}{16 \pi^2  M^2 N_c^f} \left(2 \log \frac{M}{m_{f'}}-1\right) +\nonumber \\
&\frac{|\tilde\lambda_{L/R}|^2 m_Z^2 N_c^{f'}}{288 \pi^2  M^2 N_c^f} \left\{g^f_{L/R} + g^{f'}_{R/L}\left(12 \log \frac{M}{m_{f'}}-9\right)\pm 3 T_3^{f'}\right\},
\end{align}
where $M$ is now the mass of $\tilde S$.

Note that these relations are only valid up to leading order in $m_Z/M$ and $m_{f'}/M$. Nonetheless, in Eqs.~(\ref{AppdeltagmL1}) and (\ref{AppdeltagmL2}), the terms proportional to $m_{f'}^2$ will be much larger than the ones proportional to $m_Z^2$ when the internal fermion, $f'$, is taken to be the top quark. Thus it will be these terms, proportional to $m^2_{f'}=m^2_t$, that are particularly constrained by electroweak precision measurements.

The interactions in the RPV Lagrangian in Eq.~(\ref{RPVlag}) mimic those of our toy model in Eq.~(\ref{TOYlag}). Thus, Eqs.~(\ref{Appdeltagmz1})-(\ref{AppdeltagmL2}) can also be used to compute the RPV-induced corrections to the $Z \bar f f$ couplings. To calculate the modification in the $Z \bar f f$ coupling for a fixed final state $f$, the contributions from all the possible virtual states that can couple to $f$ should be combined. For instance, a non-zero $\lambda'_{332}$ can modify the $Z\bar\tau \tau$ coupling, in addition to the $Z\bar\nu_\tau \nu_\tau$, $Z\bar b b$ and $Z\bar s s$ couplings. For the final state $\tau$, there can be two combinations of fermions and sfermions in the loop, \ie, $t_L$-quark and $\tilde s_R$, and $s_R$-quark and $\tilde t_L$. The contributions from these two virtual states can be calculated using Eqs.~(\ref{AppdeltagmL1}) and (\ref{Appdeltagmz2}) to obtain the correction to the $Z-\tau_L$ interaction: 
\begin{align}\label{deltagZtau}
	&\Delta g_{L}^{\tau} (\mbox{from }\lambda'_{332})=  \nonumber\\
	&~~~\frac{3 |\lambda'_{332}|^2 m_t^2}{32 \pi^2  m_{\tilde s_R}^2} \left(2 \log \frac{m_{\tilde s_R}}{m_t} - 1 \right)
	+\frac{ |\lambda'_{332}| ^2m_Z^2}{96\pi ^2  m_{\tilde t_L}^2} \times\nonumber\\
	& \left\{ \left(-\frac{1}{2}+s_w^2\right)+\left(\frac{1}{3} s_w^2\right) \left(12 \log\frac{m_{\tilde t_L}}{m_Z}+1+ \boldsymbol{i}6\pi\right)\right\}
\end{align}
In the above equation, we have ignored a negligible contribution proportional to $m_Z^2/m_{\tilde s_R}^2$, arising from the second term of Eq.~(\ref{AppdeltagmL1}). A similar analysis can be done for any non-zero $\lambda'_{ijk}$ in any $Z\to f\bar f$ channel.

The RPV parameter space is constrained by using $\Delta g^f$ to predict the $Z$-pole observables and comparing them with their experimental measurements. The electroweak observables used in this work are shown in Table~\ref{tab:EWTable}, where: $\Gamma(\text{inv})$ is the invisible decay width of the $Z$, $R_\ell\equiv \Gamma(\text{had})/\Gamma(\ell\bar{\ell})$, $R_q\equiv \Gamma(q \bar{q})/\Gamma(\text{had})$, and $A_f\equiv (2 g_A^f g_V^f)/((g_A^f)^2 + (g_V^f)^2)$. In these expressions, $\Gamma(\text{had})$ is the partial width of $Z$ into hadrons, and $g_V^f$ and $g_A^f$ are the effective vector and axial couplings of $Z\to f \bar f$.  

\begin{table}[tb]
	\begin{ruledtabular}
		{\renewcommand{\arraystretch}{1.5}
			\begin{tabular}{||l|ccr||}
				\textrm{Observable}&
				\textrm{Experimental}&
				\textrm{Standard~Model}&
				\textrm{Pull}\\
				\hline 
				\rule{0pt}{3ex} 
				$\Gamma(\text{inv})~[\text{MeV}]$ &    $499.0 \pm 1.5$ & $501.66 \pm 0.05$ & $-1.8$\\
				$R_e$ & $20.804 \pm 0.050$ & $20.734 \pm 0.010$ & $1.4$\\
				$R_\mu$ & $20.785 \pm 0.033$ & $20.734 \pm 0.010$ & $1.6$\\
				$R_\tau$ & $20.764 \pm 0.045$ & $20.779 \pm 0.010$ & $-0.3$\\
				$R_b$ & $0.21629 \pm 0.00066$ & $0.21579 \pm 0.00003$ & $0.8$\\
				$R_c$ & $0.1721 \pm 0.0030$ & $0.17221 \pm 0.00003$ & $0.0$\\
				$A_e$ & $0.15138 \pm 0.00216$	& $0.1470 \pm 0.0004$ & $2.0$\\
				$A_\mu$ & $0.142 \pm 0.015$	& $0.1470 \pm 0.0004$ & $-0.7$\\
				$A_\tau$ & $0.1439 \pm 0.0043$	& $0.1470 \pm 0.0004$ & $-0.7$\\
				$A_b$ & $0.923 \pm 0.020$ & $0.9347$ & $-0.6$\\
				$A_c$ & $0.670 \pm 0.027$ & $0.6678 \pm 0.0002$ & $0.1$\\
				$A_s$ & $0.895 \pm 0.091$ & $0.9356$ & $-0.4$\\
				$^\dagger g^e_V$ & $-0.03817 \pm 0.00047$ & & \\
				$^\dagger g^\mu_V$ & $-0.0367 \pm 0.0023$ & & \\
				$^\dagger g^\tau_V$ & $-0.0366 \pm 0.0010$ & & \\
				$^\dagger g^e_A$ & $-0.50111 \pm 0.00035$ & & \\
				$^\dagger g^\mu_A$ & $-0.50120 \pm 0.00054$ & & \\
				$^\dagger g^\tau_A$ & $-0.50204 \pm 0.00064$ & & \\
			\end{tabular}
		}
	\end{ruledtabular}
	\caption{\label{tab:EWTable}%
\small		The relevant LEP and SLC observables with their SM predictions~\cite{Tanabashi:2018oca}. The value of $A_\tau$ corresponds to measurements at LEP using $\tau$-lepton polarization.\\ $^{(\dagger)}$These observables are only available in the electronic version of the Review of Particle Physics.}
\end{table}

We are also using in this analysis an additional measure of lepton flavor universality that can be extracted from the electroweak data. In particular, we define four observables:
\begin{align}
	& V_{\ell e} \equiv g^\ell_V / g^e_V & A_{\ell e} \equiv g^\ell_A / g^e_A, 
\end{align}
for $\ell=\mu,\tau$. The observables $V_{\ell e}$ and $A_{\ell e}$ are measures of lepton flavor universality in the couplings of the $Z$-boson and should be unity in the SM; they were found to provide strong constraints on new sources of lepton non-universality in Ref.~\cite{Feruglio:2017rjo}. The best fit values for the $g^\ell_{V,A}$ are obtained by the Particle Data Group~\cite{Tanabashi:2018oca} but published only in the online version of the Review of Particle Physics. For the four ratios, we obtain:
\begin{align}
	 V_{\mu e}  &= 0.961 \pm 0.063 & V_{\tau e} &= 0.959 \pm 0.029 \\[1mm]
	A_{\mu e}  &= 1.0002 \pm 0.0014 & A_{\tau e} &= 1.0019 \pm 0.0015
\end{align}
where we have used the error correlation matrix for the $g^\ell_{V,A}$ from Ref.~\cite{ALEPH:2005ab} to obtain the quoted errors. We will find that the observable $A_{\tau e}$ provides the strongest current constraint on the $\lambda'_{33k}$ couplings.

We place $2\sigma$ limits on the RPV couplings by calculating changes in each observable at each point in the parameter space of ($m_{\tilde f}$, $\lambda'_{ijk}$).
The masses of sfermions, $m_{\tilde f}$, are assumed to be degenerate in these calculations. Since the SM prediction for $A_e$ is already $2\sigma$ away from the measurements (see Table~\ref{tab:EWTable}), the limits from $A_e$ are obtained at $3\sigma$ (otherwise, the whole parameter space for $\lambda'_{13k}$ is excluded by $A_e$, as these couplings always worsen the fit). We perform this analysis for each observable independently to find the strongest limit, which is then fit to a  straight line for $m_{\tilde f}\geq 1 \tev$ and quoted in Table~\ref{tab:i1}. Note that the limits we mention should not be assumed to be valid for sfermion masses below $1\tev$.

In Table~\ref{tab:i1}, the electroweak limits on $\lambda'_{3jk}$ for all $k$ are given. For the $\lambda'_{33k}$ coupling, we also show a bound from the literature. The quoted literature bound is actually an extrapolation of the bound given in the oft-cited Refs.~\cite{Allanach:1999ic,Dercks:2017lfq} and originally derived in Ref.~\cite{Bhattacharyya:1995pr}. This outdated bound is usually quoted as $\lambda'_{33k}<0.45$ for a sfermion mass of $100\gev$, and derived from $R_\tau$. Using the same data and predictions as Ref.~\cite{Bhattacharyya:1995pr}, we reproduced that result but also derived the bound for sfermion masses of 1 TeV. It is this bound that we show in Table~\ref{tab:i1}. 

Our new bounds in Table~\ref{tab:i1} come from $\Gamma(\text{inv})$ for $j=1, 2$, and from $A_{\tau e}$ for $j=3$; we find the latter provides a stronger bound than $R_\tau$. Note that, as expected, the constraints on $\lambda'_{33k}$ are much stronger than the others due to the presence of top quarks in the loops. Meanwhile, the strength of the $\Gamma(\text{inv})$ constraint is due, in large part, to the fact that the SM prediction for $\Gamma(\text{inv})$ is already higher than the experimental value by $\sim 1.8 \sigma$, and the RPV contribution always makes this discrepancy worse. 

It is useful to point out that the limits from $\Gamma(\text{inv})$ are actually the same for all $\lambda'_{ijk}$. This is because the RPV-induced $Z\to\nu\bar\nu$ process can only have down quarks in the loop (see the Lagrangian of Eq.~(\ref{RPVlag})). And since $m_{d,s,b}\ll m_Z$, the RPV-induced $\Delta g^\nu_L$ is independent of the down-quark generation index (which, in this case is both $j$ and $k$). 

Using the procedure outlined here, we also obtained the electroweak limits for all the $\lambda'_{ijk}$ couplings. We find the strongest electroweak constraint on each of the $\lambda'_{ijk}$ couplings to be as follows:
\begin{align}
\text{For } i=&1,2,3;~j=1,2;~k=1,2,3: \nonumber\\
\label{Zi12k}& \lambda'_{i1k},\lambda'_{i2k}<1.5\frac{m_{\tilde f}}{1\tev}+0.41 ~~\text{from}~\Gamma(\text{inv})
\end{align}
\begin{align}
\text{For } i=&1;~j=3;~k=1,2,3: \nonumber\\
\label{Z13k}& \lambda'_{13k}<0.51\frac{m_{\tilde f}}{1\tev}+0.36~~ \text{from $A_e$ at $3\sigma$}
\end{align}
\begin{align}
\text{For } i=&2;~j=3;~k=1,2,3:  \nonumber\\
\label{Z23k}& \lambda'_{23k}<0.66\frac{m_{\tilde f}}{1\tev}+0.42 ~~ \text{from}~R_\mu
\end{align}
\begin{align}
\text{For } i=&3;~j=3;~k=1,2,3:  \nonumber \\
\label{Z33k}& \lambda'_{33k}<0.54\frac{m_{\tilde f}}{1\tev}+0.38 ~~ \text{from}~A_{\tau e }.
\end{align}
Here, $m_{\tilde f}$ is the mass of sfermions, taken to be degenerate and $\geq1\tev$. 

All of these bounds for $i=1,2$, with two exceptions, are weaker than those we previously obtained from the LHC data in Ref.~\cite{Bansal:2018dge}. The two exceptions to this statement are: $\lambda'_{132}$ (a $3\sigma$ bound from $A_e$), which is about 10\% stronger than the bound obtained from the DY data, and is even stronger than our projected bound for the high-luminosity LHC; and $\lambda'_{232}$ (obtained from $R_\mu$), which is a few percent stronger than our current bound, and is similar to the expected bound from the high-luminosity LHC. Therefore, to reiterate, we find that:
\begin{align}
\lambda'_{132}&<0.51\frac{m_{\tilde f}}{1\tev}+0.36 \\[1mm]
\lambda'_{232}&<0.66\frac{m_{\tilde f}}{1\tev}+0.42
\end{align}
replace the bounds from DY data in Ref.~\cite{Bansal:2018dge} as the strongest currently available bounds on these two couplings.

It has been noted by previous studies that bounds on certain of the $\lambda'_{ijk}$ derived by demanding that the renormalization group running of those couplings, along with the top quark Yukawa coupling, remain perturbative up to a scale around $10^{16}\gev$; these ``perturbativity" bounds are calculated assuming a minimal low-energy particle content, a ``grand desert" and no significant high-scale threshold effects. The  resulting bounds on $\lambda'_{ijk}$ are typically between $1.0$ - $1.1$ for sfermion masses around $100\gev$ and are only slightly weakened for masses between $1$ - $10\tev$. In previous analyses~\cite{Allanach:1999ic}, the perturbativity bounds were stronger than the experimental bounds for $\lambda'_{221}$ and $\lambda'_{3jk}$ (all $j,k$). However, this analysis shows that, for sfermion masses of $1\tev$, constraints from the LHC outperform the perturbativity constraints for all $\lambda'_{31k}$, and that electroweak precision constraints outperform perturbativity constraints for $\lambda'_{33k}$; in our previous paper~\cite{Bansal:2018dge} we also found much stronger constraints from the LHC for $\lambda'_{221}$. Therefore, only for the three couplings $\lambda'_{32k}$ does one find that the perturbativity constraint (namely, $\lambda'_{32k}<1.1$) remains stronger than the experimental constraints for $m_{\tilde f}=1\tev$; we do not anticipate this situation changing with additional LHC data without a significant increase in $\tau$ tagging efficiency.

\section{\label{section5} Conclusions}
In this paper, we have first obtained limits on the RPV couplings $\lambda'_{3jk}$ using Drell-Yan $pp\to \tau\nu$ data from the LHC. We find that the LHC data can be used to derive constraints on six of the nine $\lambda'_{3jk}$ couplings that are stronger than any previously obtained, and valid for down-type squark masses above 1 TeV. We also have reanalyzed the constraints on all the $\lambda'_{ijk}$ couplings from the precision electroweak data, updating previous bounds in the literature. In particular, we found that the three $\lambda'_{33k}$ couplings are still best constrained by electroweak data, but that the constraints are now roughly 25\% stronger than previously reported, most of that improvement coming from using the lepton-flavor-violating observable $A_{\tau e}$. 

Taking both the LHC data and the precision electroweak fits into account, we have obtained new, stronger bounds on the $\lambda'_{3jk}$ RPV couplings that are valid in the TeV mass range. While we do not expect significant improvements to be forthcoming in the electroweak fits, the analysis of dilepton DY data at the LHC ($pp\to\ell^+\ell^-$) could significantly improve some of the bounds given here. And because our LHC-derived bounds come from the exchange of squarks in the $t$-channel, rather than from on-shell pair production, strengthening these bounds does not necessarily require additional center-of-mass energy, but will happen automatically with the additional luminosity one expects in the next phases of the LHC's experimental program.

~

\section*{Acknowledgments}

This work was partially supported by the National Science Foundation under grant PHY-1820860.  The work of MQ is partly supported by Spanish MINEICO (grants CICYT-FEDER-FPA2014-55613-P and FPA2017-88915-P), by the Catalan Government under grant 2017SGR1069, and Severo Ochoa Excellence Program of MINEICO (grant SEV-2016-0588).

\section*{Appendix}

In this Appendix, we pull together all of the updated limits on the 27 $\lambda'_{ijk}$ couplings for completeness. The bounds are given in Table~\ref{tab:app}. For each choice of $\{ijk\}$, the table shows the currently strongest bound on that coupling. We also show the approximate sfermion mass bound (in TeV) one obtains assuming $\lambda'_{ijk}=1$. 

\begin{table*}[t]
	\begin{ruledtabular}
		{\renewcommand{\arraystretch}{2.5}
			\begin{tabular}{||c| c c c c||}
				& &  & Sfermion Mass Limit  & \\[-5mm]
				$ijk$ & Current Bound & Source & (TeV) & \\
				\hline \hline
				\ldelim\{{2}{5mm}[111~~] & $0.16 \frac{m_{\tilde{d}_R}}{1 \tev}+0.030$ & DY monolepton & $6.1$ & \\ 
				       & $0.16 \left(\frac{m_{\mathrm{SUSY}}}{1\tev}\right)^{5/2}$ & $0\nu\beta\beta$ & $2.1$ & \\
				112, 113 &  $0.16 \frac{m_{\tilde{d}_R}}{1 \tev}+0.030$ & DY monolepton & $6.1$ & \\ 
				
				 \ldelim\{{2}{5mm}[121~~]& $0.34 \frac{m_{\tilde{q}}}{1 \tev}+0.18$ & DY dilepton  & $2.4$ & \\ 
				       & $0.43  \frac{m_{\tilde{s}_R}}{1 \tev}$ & CC universality & $2.3$ & \\ 
				122 & $0.076  \sqrt{\frac{m_{\tilde{s}}}{1 \tev}}$ & ${\nu_e}$ mass & $173$ & \\ 
				123 & $0.43  \frac{m_{\tilde{s}_R}}{1 \tev}$ & CC universality & $2.3$ & \\ 
								
				131 & $0.19  \frac{m_{\tilde{t}_L}}{1 \tev}$ & APV & $5.3$ & \\ 
				132 & $0.51 \frac{m_{\tilde{q}}}{1 \tev}+0.36$ & $A_e (3\sigma)^{~\dagger}$ & $1.3$ &  \\  
				133 & $0.0017  \sqrt{\frac{m_{\tilde{b}}}{1 \tev}}$ & ${\nu_e}$ mass & $3.5\times10^5$ & \\
			
				211, 212, 213 & $0.090 \frac{m_{\tilde{d}_R}}{1 \tev}+0.014$ & DY monolepton  & $11$ & \\ 
				
				221 & $0.34 \frac{m_{\tilde{q}}}{1 \tev}+0.074$ & DY dilepton  & $2.7$ & \\ 
				222, 223 & $0.44 \frac{m_{\tilde{s}_R}}{1 \tev}+0.040$ & DY monolepton  & $2.2$ & \\ 
				
				231 & $0.34 \frac{m_{\tilde{q}}}{1 \tev}+0.074$ & DY dilepton  & $2.7$ &  \\ 
				232 & $0.66 \frac{m_{\tilde{q}}}{1 \tev}+0.42$ & $R_\mu\,^{\dagger}$ & $<1$ &  \\ 
				233 & $0.51  \sqrt{\frac{m_{\tilde{b}}}{1 \tev}}$ & ${\nu_\mu}$ mass & $3.8$ &  \\
			
				311, 312, 313 & $0.20 \frac{m_{\tilde{d}_R^k}}{1 \tev}+0.046$ & DY monolepton$^{~\dagger}$ & $4.8$ &  \\ 
					
				321, 322, 323 & $1.5 \frac{m_{\tilde{d}_R^k}}{1 \tev}+0.66$ & DY monolepton$^{~\dagger}$ & $<1$ & \\ 
				
				331, 332, 333 & $0.54 \frac{m_{\tilde{f}}}{1 \tev}+0.38$ & $A_{\tau e}\,^{\dagger}$ & $1.1$ & 

			\end{tabular}
		}
	\end{ruledtabular}
	\caption{\small \label{tab:app}%
		Current upper bounds on $\lambda'_{ijk}$, the source of each bound, and the excluded sfermion masses assuming the corresponding $\lambda'_{ijk}=1$. Bounds obtained in this work are indicated with a dagger ($^\dagger$). For the 111 and 121 entries, more than one bound compete for masses above $1\tev$.}
\end{table*}

Entries in which the best limits are derived in this paper are indicated with a dagger. Other bounds come from a variety of processes: 
\begin{itemize}
\item $11k, 121, 21k, 22k, 231$ are bounded from mono- and dilepton DY scattering into final state electrons and muons, as derived in Ref.~\cite{Bansal:2018dge}; 
\item $111$ has an additional bound from neutrinoless double beta decay ($0\nu\beta\beta$), but which falls quickly with the mass scale of the SUSY particles (specifically, the selectron and neutralino), taken to be degenerate here at $m_{\mathrm{SUSY}}$~\cite{Mohapatra:1986su,Allanach:1999ic};
\item $122, 123$ bounds are from charged current universality in $\beta$-decay, and 131 is from atomic parity violation (APV), all of which were first derived in Ref.~\cite{Barger:1989rk} and updated in Refs.~\cite{ledroit:in2p3,Allanach:1999ic}; 
\item $133, 233$ bounds come from constraints on contributions to neutrino masses, first derived in Refs.~\cite{Mohapatra:1986su,Godbole:1992fb} and updated here with the current PDG bounds $m_{\nu_e}<2\,$eV and $m_{\nu_\mu}<0.17\,$MeV~\cite{Tanabashi:2018oca}. These update the bounds listed in Ref.~\cite{Bansal:2018dge}.
\end{itemize}
One sees from the table that Drell-Yan scattering at the LHC already provides the strongest bounds on 17 of the 27 coupling constants. With additional luminosity, the LHC is capable of obtaining the strongest bound on at least one additional coupling (\ie, 131, see~\cite{Bansal:2018dge} for projections for the reach of a high-luminosity LHC for the $1jk$ and $2jk$ couplings).


\begin{thebibliography}{99}

\bibitem{Canepa:2019hph} 
  For a recent review of the status of LHC searches for supersymmetry, see: A.~Canepa,
  %``Searches for Supersymmetry at the Large Hadron Collider,''
  Rev.\ Phys.\  {\bf 4}, 100033 (2019).
 % doi:10.1016/j.revip.2019.100033
  %%CITATION = doi:10.1016/j.revip.2019.100033;%%

\bibitem{Bansal:2018dge} 
 S.~Bansal, A.~Delgado, C.~Kolda and M.~Quiros,
 %``Limits on R-parity-violating couplings from Drell-Yan processes at the LHC,''
 arXiv:1812.04232 [hep-ph].
 %%CITATION = ARXIV:1812.04232;%%
 
\bibitem{Raj:2016aky} 
 N.~Raj,
 %``Anticipating nonresonant new physics in dilepton angular spectra at the LHC,''
 Phys.\ Rev.\ D {\bf 95}, 015011 (2017)
 %doi:10.1103/PhysRevD.95.015011
 [arXiv:1610.03795 [hep-ph]].
 %%CITATION = doi:10.1103/PhysRevD.95.015011;%%
 
 \bibitem{Alves:2018krf} 
  A.~Alves, O.~J.~P.~Eboli, G.~Grilli Di Cortona and R.~R.~Moreira,
  %``Indirect and monojet constraints on scalar leptoquarks,''
  Phys.\ Rev.\ D {\bf 99}, no. 9, 095005 (2019)
  %doi:10.1103/PhysRevD.99.095005
  [arXiv:1812.08632 [hep-ph]].
  %%CITATION = doi:10.1103/PhysRevD.99.095005;%%
 
 \bibitem{Bansal:2018eha} 
 S.~Bansal, R.~M.~Capdevilla, A.~Delgado, C.~Kolda, A.~Martin and N.~Raj,
 %``Hunting leptoquarks in monolepton searches,''
 Phys.\ Rev.\ D {\bf 98}, 015037 (2018)
 %doi:10.1103/PhysRevD.98.015037
 [arXiv:1806.02370 [hep-ph]].
 %%CITATION = doi:10.1103/PhysRevD.98.015037;%%
  
\bibitem{Aaboud:2018vgh} 
M.~Aaboud {\it et al.} [ATLAS Collaboration],
%``Search for High-Mass Resonances Decaying to $\tau\nu$ in pp Collisions at $\sqrt{s}$=13  TeV with the ATLAS Detector,''
Phys.\ Rev.\ Lett.\  {\bf 120}, no. 16, 161802 (2018)
%doi:10.1103/PhysRevLett.120.161802
[arXiv:1801.06992 [hep-ex]].
%%CITATION = doi:10.1103/PhysRevLett.120.161802;%%

\bibitem{Sirunyan:2018lbg} 
A.~M.~Sirunyan {\it et al.} [CMS Collaboration],
%``Search for a W' boson decaying to a $\tau$ lepton and a neutrino in proton-proton collisions at $\sqrt{s} =$ 13 TeV,''
Phys.\ Lett.\ B {\bf 792}, 107 (2019)
%doi:10.1016/j.physletb.2019.01.069
[arXiv:1807.11421 [hep-ex]].
%%CITATION = doi:10.1016/j.physletb.2019.01.069;%%
 
 \bibitem{Sirunyan:2018qio} 
 A.~M.~Sirunyan {\it et al.} [CMS Collaboration],
 %``Measurement of the $\mathrm{Z}\gamma^{*} \to \tau\tau$ cross section in pp collisions at $\sqrt{s} = $ 13 TeV and validation of $\tau$ lepton analysis techniques,''
 Eur.\ Phys.\ J.\ C {\bf 78}, no. 9, 708 (2018)
 %doi:10.1140/epjc/s10052-018-6146-9
 [arXiv:1801.03535 [hep-ex]].
 %%CITATION = doi:10.1140/epjc/s10052-018-6146-9;%%

 \bibitem{Martin:2009iq} 
 A.~D.~Martin, W.~J.~Stirling, R.~S.~Thorne and G.~Watt,
 %``Parton distributions for the LHC,''
 Eur.\ Phys.\ J.\ C {\bf 63}, 189 (2009)
 %doi:10.1140/epjc/s10052-009-1072-5
 [arXiv:0901.0002 [hep-ph]].
 %%CITATION = doi:10.1140/epjc/s10052-009-1072-5;%%

\bibitem{Barger:1989rk} 
  V.~D.~Barger, G.~F.~Giudice and T.~Han,
  %``Some New Aspects of Supersymmetry R-Parity Violating Interactions,''
  Phys.\ Rev.\ D {\bf 40}, 2987 (1989).
  %doi:10.1103/PhysRevD.40.2987
  %%CITATION = doi:10.1103/PhysRevD.40.2987;%%
    
\bibitem{ledroit:in2p3}
 Ledroit, F. and Sajot, G.,
 %``Indirect limits on SUSY Rp violating couplings lambda and lambda,"
 in2p3-00362621 (1998) http://hal.in2p3.fr/in2p3-00362621
 

\bibitem{Allanach:1999ic} 
  B.~C.~Allanach, A.~Dedes and H.~K.~Dreiner,
  %``Bounds on R-parity violating couplings at the weak scale and at the GUT scale,''
  Phys.\ Rev.\ D {\bf 60}, 075014 (1999)
  %doi:10.1103/PhysRevD.60.075014
  [hep-ph/9906209].
  %%CITATION = doi:10.1103/PhysRevD.60.075014;%%
  
\bibitem{Barbier:2004ez} 
  R.~Barbier {\it et al.},
  %``R-parity violating supersymmetry,''
  Phys.\ Rept.\  {\bf 420}, 1 (2005)
  %doi:10.1016/j.physrep.2005.08.006
  [hep-ph/0406039].
  %%CITATION = doi:10.1016/j.physrep.2005.08.006;%%
  
\bibitem{Chemtob:2004xr} 
  M.~Chemtob,
  %``Phenomenological constraints on broken R parity symmetry in supersymmetry models,''
  Prog.\ Part.\ Nucl.\ Phys.\  {\bf 54}, 71 (2005)
  %doi:10.1016/j.ppnp.2004.06.001
  [hep-ph/0406029].
  %%CITATION = doi:10.1016/j.ppnp.2004.06.001;%%
  
 \bibitem{Dercks:2017lfq} 
  D.~Dercks, H.~Dreiner, M.~E.~Krauss, T.~Opferkuch and A.~Reinert,
  %``R-Parity Violation at the LHC,''
  Eur.\ Phys.\ J.\ C {\bf 77}, no. 12, 856 (2017)
  %doi:10.1140/epjc/s10052-017-5414-4
  [arXiv:1706.09418 [hep-ph]].
  %%CITATION = doi:10.1140/epjc/s10052-017-5414-4;%%
  
 \bibitem{Mizukoshi:1994zy} 
J.~K.~Mizukoshi, O.~J.~P.~Eboli and M.~C.~Gonzalez-Garcia,
%``Bounds on scalar leptoquarks from $Z$ physics,''
Nucl.\ Phys.\ B {\bf 443}, 20 (1995)
%doi:10.1016/0550-3213(95)00162-L
[hep-ph/9411392].
%%CITATION = doi:10.1016/0550-3213(95)00162-L;%%

\bibitem{ColuccioLeskow:2016dox} 
E.~Coluccio Leskow, G.~D'Ambrosio, A.~Crivellin and D.~Müller,
%``$(g-2)_\mu$, lepton flavor violation, and $Z$ decays with leptoquarks: Correlations and future prospects,''
Phys.\ Rev.\ D {\bf 95}, no. 5, 055018 (2017)
%doi:10.1103/PhysRevD.95.055018
[arXiv:1612.06858 [hep-ph]].
%%CITATION = doi:10.1103/PhysRevD.95.055018;%%

\bibitem{Bansal:2018nwp} 
S.~Bansal, R.~M.~Capdevilla and C.~Kolda,
%``Constraining the minimal flavor violating leptoquark explanation of the $R_{D^{(*)}}$  anomaly,''
Phys.\ Rev.\ D {\bf 99}, no. 3, 035047 (2019)
%doi:10.1103/PhysRevD.99.035047
[arXiv:1810.11588 [hep-ph]].
%%CITATION = doi:10.1103/PhysRevD.99.035047;%%

\bibitem{Godbole:1992fb} 
R.~M.~Godbole, P.~Roy and X.~Tata,
%``Tau signals of R-parity breaking at LEP-200,''
Nucl.\ Phys.\ B {\bf 401}, 67 (1993)
%doi:10.1016/0550-3213(93)90298-4
[hep-ph/9209251].
%%CITATION = doi:10.1016/0550-3213(93)90298-4;%%

\bibitem{Bhattacharyya:1995pr} 
G.~Bhattacharyya, J.~R.~Ellis and K.~Sridhar,
%``New LEP constraints on some supersymmetric Yukawa interactions that violate R-parity,''
Mod.\ Phys.\ Lett.\ A {\bf 10}, 1583 (1995)
%doi:10.1142/S0217732395001708
[hep-ph/9503264].
%%CITATION = doi:10.1142/S0217732395001708;%%

\bibitem{Yang:1999ms} 
J.~M.~Yang,
%``R(b) and R(l) in MSSM without R-parity,''
Eur.\ Phys.\ J.\ C {\bf 20}, 553 (2001)
%doi:10.1007/s100520100691
[hep-ph/9905486].
%%CITATION = doi:10.1007/s100520100691;%%

\bibitem{Lebedev:1999vc} 
O.~Lebedev, W.~Loinaz and T.~Takeuchi,
%``Constraints on R-parity violating couplings from lepton universality,''
Phys.\ Rev.\ D {\bf 61}, 115005 (2000)
%doi:10.1103/PhysRevD.61.115005
[hep-ph/9910435].
%%CITATION = doi:10.1103/PhysRevD.61.115005;%%

\bibitem{Lebedev:1999ze} 
O.~Lebedev, W.~Loinaz and T.~Takeuchi,
%``Constraints on R-parity violating couplings from LEP / SLD hadronic observables,''
Phys.\ Rev.\ D {\bf 62}, 015003 (2000)
%doi:10.1103/PhysRevD.62.015003
[hep-ph/9911479].
%%CITATION = doi:10.1103/PhysRevD.62.015003;%%

\bibitem{Tanabashi:2018oca} 
M.~Tanabashi {\it et al.} [Particle Data Group],
%``Review of Particle Physics,''
Phys.\ Rev.\ D {\bf 98}, no. 3, 030001 (2018).
%doi:10.1103/PhysRevD.98.030001
%%CITATION = doi:10.1103/PhysRevD.98.030001;%%

\bibitem{Feruglio:2017rjo} 
  F.~Feruglio, P.~Paradisi and A.~Pattori,
  %``On the Importance of Electroweak Corrections for B Anomalies,''
  JHEP {\bf 1709}, 061 (2017)
 % doi:10.1007/JHEP09(2017)061
  [arXiv:1705.00929 [hep-ph]].
  %%CITATION = doi:10.1007/JHEP09(2017)061;%%
  
%\cite{ALEPH:2005ab}
\bibitem{ALEPH:2005ab} 
  S.~Schael {\it et al.} [ALEPH and DELPHI and L3 and OPAL and SLD Collaborations and LEP Electroweak Working Group and SLD Electroweak Group and SLD Heavy Flavour Group],
  %``Precision electroweak measurements on the $Z$ resonance,''
  Phys.\ Rept.\  {\bf 427}, 257 (2006)
 % doi:10.1016/j.physrep.2005.12.006
  [hep-ex/0509008].
  %%CITATION = doi:10.1016/j.physrep.2005.12.006;%%
  
  \bibitem{Mohapatra:1986su} 
  R.~N.~Mohapatra,
  %``New Contributions to Neutrinoless Double beta Decay in Supersymmetric Theories,''
  Phys.\ Rev.\ D {\bf 34}, 3457 (1986).
  %doi:10.1103/PhysRevD.34.3457
  %%CITATION = doi:10.1103/PhysRevD.34.3457;%%

  
\end{thebibliography}
\end{document}